# Linear viscoelasticity of entangled wormlike micelles bridged by telechelic polymers : an experimental model for a double transient network

Kaori Nakaya[1,2], Laurence Ramos[1], Hervé Tabuteau[1] and Christian Ligoure[1,a]

[1] *Laboratoire des Colloïdes, Verres et Nanomatériaux, UMR CNRS/UM2 5587, Université Montpellier 2, Place Eugène Bataillon, F-34095 Montpellier Cedex 05, France*
[2] *Department of Physics, Faculty of Sciences, Ochanomizu University, 2-1-1 Otsuka, Bunkyo-ku, Tokyo 112-0012, Japan*

a) corresponding author: email: christian.ligoure@univ-montp2.fr

**SYNOPSIS**

We survey the linear viscoelasticity of a new type of transient network: bridged wormlike micelles, whose structure has been characterized recently [Ramos and Ligoure, (2007)]. This composite material is obtained by adding telechelic copolymers (water-soluble chains with hydrophobic stickers at each extremity) to a solution of entangled wormlike micelles (WM). For comparison, naked WM and WM decorated by amphiphilic copolymers are also investigated. While these latter systems exhibit almost a same single ideal Maxwell behavior, solutions of bridged WM can be described as two Maxwell fluids components blends, characterized by two markedly different characteristic times, $\tau_{fast}$ and $\tau_{slow}$, and two elastic moduli, $G_{fast}$ and $G_{slow}$, with $G_{fast} \gg G_{slow}$. We show that the slow mode is related to the viscoelasticity of the transient network of entangled WM, and the fast mode to the network of telechelic active chains (i.e. chains that do not form loops but bridge two micelles). The dependence of the viscoelasticity with the surfactant concentration, $\phi$, and the sticker-to-surfactant molar ratio, $\beta$, is discussed. In particular, we show that $G_{fast}$ is proportional to the number of active chains in the material, $\phi\beta$. Simple theoretical expectations allow then to evaluate the bridges/loops ratio for the telechelic polymers.

# Introduction

Aqueous transient self-assembled networks constitute a class of complex materials that spontaneously form *reversible* equilibrium 3-D networks in water, which are transiently capable of transmitting elastic forces over macroscopic distances. Among them, two systems have been the subject of numerous fundamental studies in the last three decades. These are entangled solutions of long and flexible surfactant cylinders (wormlike micelles, WM) on the one hand, and solutions of associated water-soluble polymers with two hydrophobic end blocks (telechelic chains) above the percolation concentration, on the other hand. Both systems display the same simple viscoelastic Maxwell behavior in the linear regime where the frequency dependence of the real $G'$ and imaginary $G''$ parts of the complex viscoelastic modulus are given by:

$$G'(\omega) = G_\infty \frac{\tau_M^2 \omega^2}{1 + \tau_M^2 \omega^2} \quad (1)$$

$$G''(\omega) = G_\infty \frac{\tau_M \omega}{1 + \tau_M^2 \omega^2} \quad (2)$$

where $\omega$ is the oscillation frequency.

For telechelic polymer solutions, the plateau modulus is given by $G_\infty = \mu_b k_B T$, where $\mu_b$ is the number of elastically active chains in the network per unit volume, $k_B$, the Boltzmann constant and $T$ the absolute temperature. The relaxation time, $\tau_M$, is the average lifetime of a link, which is in turn related to the thermally activated kinetics of detachment and reattachment of a sticker to the hydrophobic core of a micelle. Theory for the viscoelasticity of this type of transient network was developed independently by Tanaka and Edwards, (1992) and Jenkins, *et al.*, (1991) on the basis of previous works by Green and Tobolsky, (1946) and Yamamoto [Yamamoto, (1956), Yamamoto, (1957)]. In a seminal paper, Annable, *et al.*, (1993) showed that the transient network theory well describes the linear rheological behavior of solutions of telechelic polymers, with however a slight modification to take into account the presence of loops that do not contribute to the elasticity of the network. An alternative modification of the transient network theory, the so-called micellar theory, has been recently proposed by Meng and Russel, (2006).

Several experimental reports, reviewed recently by Berret, *et al.*, (2003), confirmed the validity of the transient network theory for telechelic polymers with fluorocarbon endcaps. In particular, it has been shown that the simple Maxwell behavior can be attributed to the dissociation of a hydrophobe to the junction points [Winnik and Yekta, (1997)]. Slight deviations with respect to the ideal Maxwell case may exist; the possible origin of these deviations have been elucidated by Hough and Ou-Yang, (2006). Note that a system of oil in water microemulsion droplets linked by telechelic polymers exhibits the same simple Maxwell behavior in the linear regime above some percolation threshold, in agreement with the theory for transient network [Filali, *et al.*, (2001)].

For entangled wormlike micelles (WM), the plateau modulus is related to the correlation length $\xi$ of semi-dilute polymer solutions according to $G_\infty \approx k_B T/\xi^3$. The reptation theory of breakable chains by Cates, (1987) predicts that the relaxation time is given by $\tau_M \approx \sqrt{\tau_{break}\tau_{rep}}$ for a sufficiently long breaking time ($\tau_{break} \gg \tau_{rep}$), where $\tau_{break}$ is the mean life time of a chain before breaking into two pieces and $\tau_{rep}$ is the mean reptation time of a chain. Static and dynamic behavior of surfactant WM has been reviewed by Cates and Candau, (1990). The predictions of the Cates theory were confirmed for several experimental surfactant systems of WM for which the micelles do not form junctions [Candau, *et al.*, (1988)], [Berret, *et al.*, (1993)], [Cappelaere, *et al.*, (1995)].

The aim of the present paper is to investigate a simple experimental situation of a *double transient network* obtained by mixing together two binary solutions of self assembled systems, that can form by themselves transient networks in water. Mixtures of surfactant wormlike micelles and water soluble side-chain associative polymers fall *a priori* into this category. Rheology experiments on this type of systems have demonstrated that the polymers considerably modify the viscoelasticity of the samples and important synergetic effects have been evidenced [Panmai, *et al.*, (1999)], [Couillet, *et al.*, (2005)], [Shashkina, *et al.*, (2005)]. In particular, it has been shown that the viscosity of the mixtures can be up to several orders of magnitude higher than the viscosities of the pure component solutions. However, the rheological behavior of the associative networks formed in binary solutions of side-chain associative polymers is more complex and less understood than that of network formed with telechelic chains. This renders the modeling of composite WM / water soluble side-chain associative polymers systems rather difficult. We therefore investigate an *a priori* simpler experimental double transient network, obtained by adding in a solution of entangled surfactant wormlike micelles, telechelic triblock copolymers whose hydrophobic ends anchor into the micelles and whose hydrophilic tails are swollen in the aqueous solvent and can reversibly link the micelles. For comparison we have also studied the same surfactant system decorated with an amphiphilic diblock copolymer which corresponds exactly to a triblock telechelic copolymer cut into two identical diblock copolymers. The comparison of the linear rheological behavior of both systems will allow one to emphasize the contribution of bridges between micelles to the rheological properties of the network, since a conformation entropy of loop with polymerization index $2N$ is approximately equivalent to the conformation entropy of two dangling chains of polymerization index $N$.

The outline of the paper is as follows. In section 1, we describe the materials and methods and summarize the main results obtained by two of us [Ramos and Ligoure, (2007)] on the phase behavior and structural properties of the transient network obtained by mixing WM and telechelic polymers. Section 2 reports the results for the linear rheology of double transient networks for various sticker to surfactant molar ratio $\beta$ and surfactant concentration $\phi$. In Section 3, we analyze and discuss the experimental data, and finally conclude.

## 1 Materials and methods

**Materials**

We use surfactant solutions composed of a mixture of cetylpyridinium chloride $[H_3C-(CH_2)_{15}]-C_5H_4N^+-Cl^-$ (CpCl) and sodium salicylate (NaSal), with a constant molar ratio [NaSal]/[CpCl] = 0.5, diluted in brine [NaCl]=0.5 M. CpCl is received from Fluka and is purified by successive recrystallization in water and acetone. NaSal and NaCl are used as received. This system is known to form long and flexible micelles even at low concentration as shown by Rehage and Hoffmann, (1988); its linear rheological properties have been characterized in detail by Berret*, et al.*, (1993) We add to this host phase di- or triblock copolymers, which are synthesized in our laboratory. The water soluble block is poly(ethyleneoxide) (PEO) and it has been hydrophobically modified and purified using the method described in Kaczmarski and Glass, (1993) and Vorobyova*, et al.*, (1998). Two classes of copolymers have been prepared: a triblock "telechelic" polymers, $C_{18}$-$PEO_{10K}$-$C_{18}$ with a $C_{18}H_{37}$ aliphatic chain grafted at each extremity of the central PEO chain of molecular weight 10 000 g/mol, and a diblock "amphiphilic" $PEO_{5K}$-$C_{18}$ copolymer, constituted of a polyethylene oxide (PEO) block, of molecular weight 5 000 g/mol, grafted at one extremity with a $C_{18}H_{37}$ aliphatic chain. Hence, the amphiphilic copolymer corresponds exactly to the triblock telechelic copolymer cut into two identical diblock copolymers. After modification, the degrees of substitution of the hydroxyl groups were determined by NMR using the method described by Hartmann, *et al.*, (1999) and are found to be larger than 98%. The radius of gyration of the POE block are respectively 37 Å and 24 Å [Cabane and Duplessix, (1982)] for the telechelic and amphiphilic polymer respectively. The samples are characterized by the mass fraction of surfactant $\phi = (m_{CPCl} + m_{Sal})/m_{tot}$, where $m_{CPCl}$, $m_{Sal}$, $m_{tot}$ are respectively the mass of CpCl, the mass of salicylate and the total mass of the sample, and by the sticker ($C_{18}H_{37}$) over surfactant molar ratio $\beta$. Hence for given $\phi$ and $\beta$, a mixture with di-block copolymers comprise twice as many copolymer molecules than one with telechelic polymers, but the total number of ethylene oxide units is the same. The samples are prepared by weight. We first incorporate the surfactant CpCl and the hydrophobically modified PEO in brine until complete dissolution of the polymer (this requires one night or more). After addition of NaSal to the homogeneous mixture, the sample is stirred several times for homogenization and then left undisturbed at 30°C for several days. In the following, since the densities of all components are nearly identical, we identify mass and volume fractions. Unless stated, experiments are performed at a temperature $T$ of 30°C.

**Summary of the structural results**

We summarize in this paragraph the main results on the structure and phase behavior of the mixed system as reported by Ramos and Ligoure, (2007). Without copolymer, the micellar solutions are homogeneous and isotropic for surfactant weight fraction up to 36%. (The overlap concentration of naked micelles is $\phi^* = 0.3\%$.) Addition of telechelic polymers induces an effective attraction of entropic origin between the surfactant cylinders according to Zilman and Safran, (2003) that may result, in equilibrium, in the coexistence of a dilute phase and a isotropic stiff gel, as shown in the phase diagram of the mixed system (Figure 1). By contrast, the phase diagram of naked micelles is not modified by the addition of amphiphilic copolymer (one sticker): the mixture remains a transparent isotropic phase.

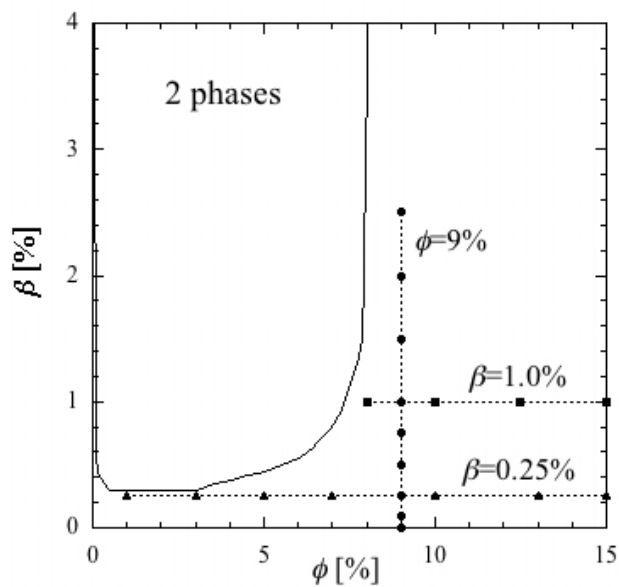

Figure. 1: Phase diagram in the plane ($\phi,\beta$), with $\phi$ the surfactant concentration, and $\beta$ the sticker over surfactant molar ratio at a temperature $T= 30°C$. Rheological measurements were done along three lines, with $\beta = 0 - 2.5\%$ at $\phi=9\%$, $\phi = 1 - 15\%$ at $\beta = 0.25\%$ and $\phi = 8 -15\%$ at $\beta = 1\%$. The symbols show the composition of the samples investigated.

Small-angle neutron scattering (SANS) experiments show first that the cylindrical structure of the micelles is maintained upon copolymer addition. On the other hand, the addition of telechelic polymer correlates with (i) the emergence of a broad peak in the structure factor at a finite scattering vector, $q$, which is also observed at the same position for solution hairy wormlike micelles with an equivalent composition (same $\phi$ and $\beta$); this peak originates from the intermicellar short range steric repulsion induced by the polymer layers, and (ii) the large rise of the scattered intensity at low scattering vectors, which is the signature of an attractive interaction between the micelles provoked by the telechelic polymers linking them. The expected (as inferred from the SANS experiments) structure of the double transient network of wormlike micelles and telechelics in the one phase domain is sketched in Figure 2: the mixed system forms a double transient network of entangled wormlike micelles that are linked by sliding junctions formed by telechelic polymers in a bridge configuration and are "decorated" by telechelic chains in a loop configuration. Moreover, as shown by Brackman and Engberts, (1989), neutral polymers such a as POE do not interact (or interact very weakly) with cationic surfactants, so that one does not expect any adsorption of the water soluble chain of the telechelic on the micelles. Finally note that there is no a experimental evidence of the existence of pure micelles of telechelic polymers in the composite system.

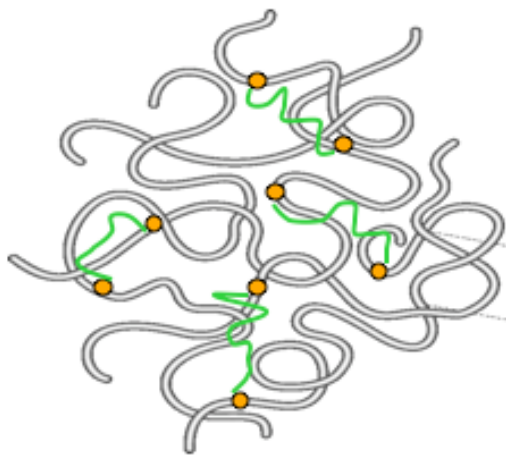

Figure 2 Sketch of the structure of the double transient network formed by addition of telechelic polymers to a solution of entangled WM in the one phase domain of the phase diagram, as inferred from structural investigations.

**Methods**

Rheological measurements are performed using a strain controlled Rheometrics fluid spectrometer (RFS II) in a Couette geometry. For each sample, a strain sweep test is performed to determine the maximum strain amplitude $\gamma$ for a linear response at two angular frequencies $\omega = 0.1$ s$^{-1}$ and $\omega = 30$ s$^{-1}$. The responses are found to be linear up to $\gamma = 30\%$ for all samples. The measurements reported here are made in the linear viscoelastic regime (LVER) with a strain amplitude in the range 5%-30% depending on the samples. Dynamical measurements are carried out for angular frequency $\omega = 0.1 - 100$ rad/s. Strain steps experiments in the LVER are systematically performed to confirm the dynamical measurements. In addition, steady shear rates measurements are made for shear rates $\dot{\gamma} = 0.2$-3 s$^{-1}$, in order to measure directly the zero shear viscosity $\eta(\dot{\gamma} \to 0)$. Rheological measurements are done along three lines of composition in the one phase domain, with $\beta = 0$ - 2.5% at $\phi = 9\%$, $\phi = 1$ - 15% at $\beta = 0.25\%$ and $\phi = 8$ - 15% at $\beta = 1\%$ (see Figure 1). For comparison the same measurements are performed for solutions of hairy wormlike micelles with equivalent compositions.

# 2 Results

**Steady-shear flow**

Figure 3 shows the Newtonian viscosity for naked wormlike micelles and mixtures of wormlike micelles with increasing amounts of amphiphilic copolymers and of telechelic copolymers, for a fixed surfactant concentration $\phi=9\%$. A slight increase of viscosity is measured in the presence of amphiphilic copolymer, as $\eta$ grows by a factor less than 3 when $\beta$ varies between 0 and 2.5%. By contrast, the viscosity increases by a factor larger than 20 when the micelles are mixed with telechelic polymers. This strong viscosity growth is a clear indication that the surfactant micelles are bridged by the telechelic block copolymers. We note that similar viscosity increase has been observed by Couillet, *et al.*, (2005), and Shashkina, *et al.*, (2005) in mixtures of surfactant wormlike micelles and polymers that bear hydrophobic stickers randomly distributed along a hydrophilic backbone.

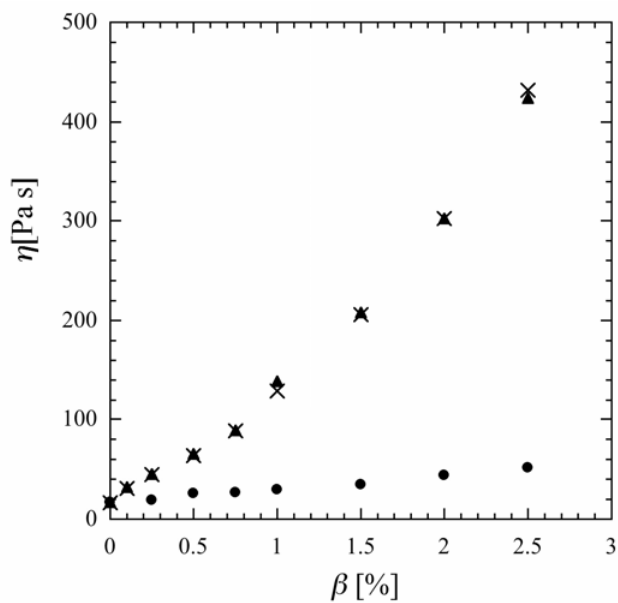

Figure. 3: Zero-shear viscosity, $\eta$, for hairy (closed circles) and bridged (closed triangles and crosses) WM at a surfactant concentration $\phi = 9\%$ as a function of the amount of polymer, $\beta$. The viscosity is measured from steady shear rate sweep tests (closed triangles) and dynamical frequency sweep tests (crosses). The temperature is $T=30°C$.

**Dynamic viscoelasticity**

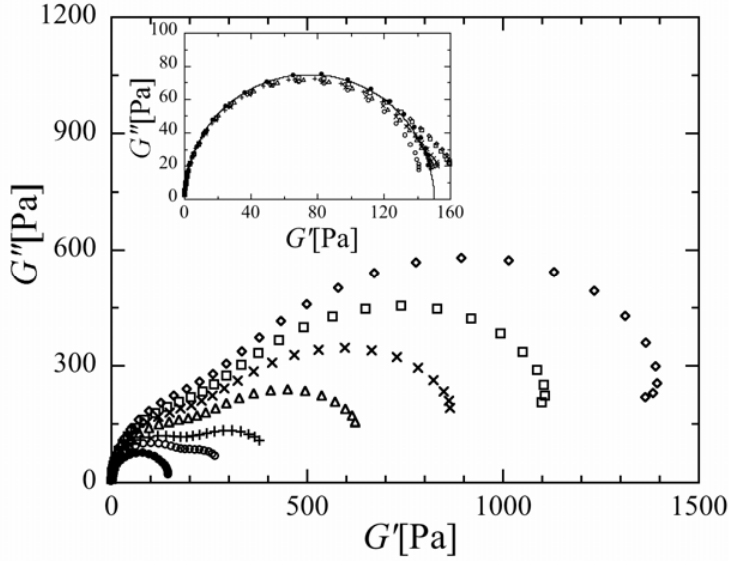

Figure 4: Cole-Cole plot for WM with various telechelic polymer concentration, $\beta$ = 0 (closed circles), 0.25% (opened circles), 0.5% (plusses), 1.0% (triangles), 1.5% (crosses), 2.0% (squares), 2,5% (diamonds). Inset: Cole-Cole plot for WM with various amphiphilic polymer concentration $\beta$ (same symbols as the main figure). In both graphs, the surfactant concentration is $\phi$ = 9% and the temperature is $T$=30°C.

To better understand the origin of the dramatic increase of viscosity for the mixture of wormlike micelles and telechelic polymers and ultimately model the viscoelasticity of this class of composite materials, we have performed dynamic experiments, in the linear regime. Figure 4 shows Cole-Cole plots, $G''$, the loss modulus, as a function of $G'$, the storage modulus, and Figures 5 show the frequency dependence of $G'$ and $G''$. Solutions of pure surfactant wormlike micelles and of wormlike micelles decorated by amphiphilic polymers exhibit a Maxwellian behavior expressed by Eqs (1) and (2), as revealed by a perfect semi-circle in the Cole-Cole plot (inset of figure 3). Indeed the applicability of Eqs (1) and (2) implies that the data should be in the form of a semicircle described by :

$$G''(\omega) = \sqrt{G'(\omega)G_\infty - G'(\omega)^2} \qquad (3)$$

Moreover, the almost perfect superposition of the data obtained with different copolymer amounts with that of pure surfactant micelles demonstrates that the addition of amphiphilic polymer does not affect significantly the viscoelasticity of a semi-dilute solution of wormlike micelles. Slight deviations of the semicircle shape are observed at high frequency, and seems to increase with increasing $\beta$; they have been attributed to a crossover towards the breathing regime in the relaxation process of wormlike micelles by Granek and Cates, (1992) and have already observed by Berret, *et al.*, (1993), and in more detail by Cardinaux, *et al.*, (2002) and Buchanan, *et al.*, (2005) using micro-rheology techniques.

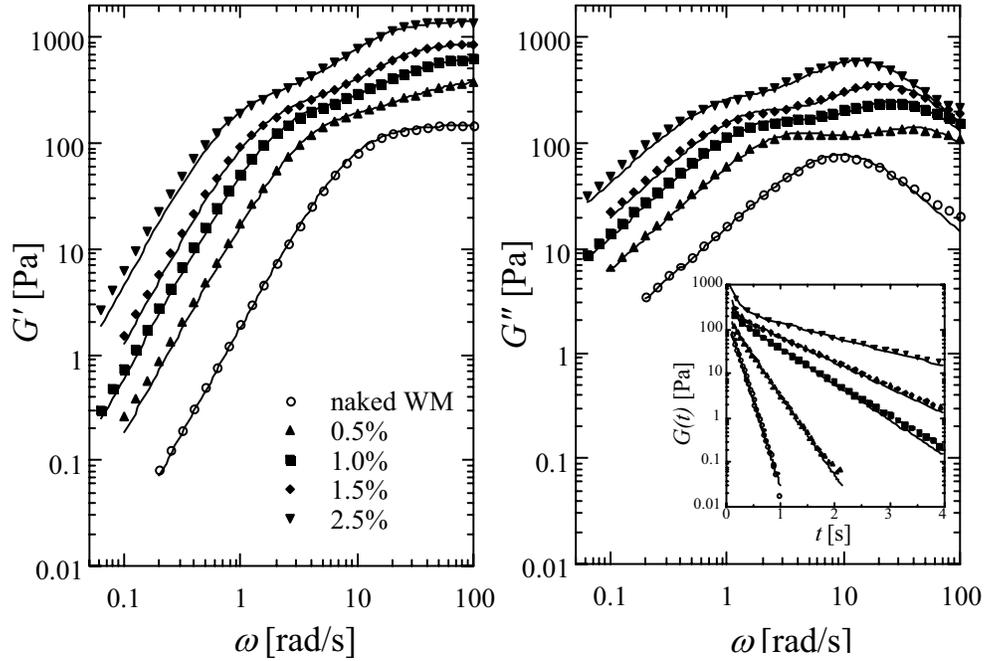

Figure 5: Storage modulus, *G'*, (left) and loss modulus, *G"*, (right) as a function of the frequency, $\omega$, for WM solutions with various amount of telechelic polymers, $\beta$, in the range 0 - 2.5%. The surfactant concentration is $\phi$ = 9% and the temperature is *T*=30˚C. Symbols are the same for the two graphs. Solid lines correspond to fits by two Maxwell components in parallel (Eqs. 4 and 5), except for naked micelles where the fits correspond to Eqs. (1,2). The inset shows the stress relaxation with naked WM and bridged WM (same symbols as the main figure). Solid lines correspond to fits with a simple exponential decay (for naked WM) and with Eq. (6) for bridged WM.

Interestingly, while the visco-elasticity of pure wormlike micelles or hairy wormlike micelles is characterized, in the range of frequency experimentally accessible, by a unique characteristic time, the visco-elasticity of wormlike micelles bridged by telechelic polymers is characterized by two distinct characteristic times. These are clearly revealed both in the Cole-Cole plots, and in the frequency dependence of the complex modulus. The Cole-Cole plots exhibit a growth of *G'* and *G"* as the amount of telechelic increases and a transition from a semi-circle to a combination of two semi-circles. Accordingly, the frequency dependence of *G'* and *G"* varies from that of a Maxwell fluid, for naked micelles, to that a visco-elastic fluid with two distinct modes, when the micelles are bridged by telechelics. Note that the occurrence of two distinct relaxation modes is particularly clearly visible in the two local maxima of the frequency dependence of the loss modulus. More quantitatively, the data for micelles connected by telechelic polymers can only be satisfyingly fitted with the expression for a two-modes Maxwell model consisting of two Maxwell elements connected in parallel, for which the frequency dependence of $G'$ and $G''$ reads:

$$G'(\omega) = G_{slow}\frac{\tau_{slow}^2\omega^2}{1+\tau_{slow}^2\omega^2} + G_{fast}\frac{\tau_{fast}^2\omega^2}{1+\tau_{fast}^2\omega^2} \qquad (4)$$

$$G''(\omega) = G_{slow}\frac{\tau_{slow}\omega}{1+\tau_{slow}^2\omega^2} + G_{fast}\frac{\tau_{fast}\omega}{1+\tau_{fast}^2\omega^2} \qquad (5)$$

In marked contrast, only one mode (Eqs(1,2)) is sufficient to fit satisfyingly the data for naked micelles, and for hairy micelles (data not shown). The fits are shown as continuous lines in Figure 5. In the inset of Figure 5b, we also show the stress relaxation data, which clearly exhibit as well a two-time relaxation process when the micelles are connected with telechelic polymers. Accordingly, the stress relaxation can be fitted with a sum of two exponential decays:

$$G(t) = G_{fast}\exp(-t/\tau_{fast}) + G_{slow}\exp(-t/\tau_{slow}) \qquad (6)$$

The four fit parameters $(G_{slow}, G_{fast}, \tau_{slow}, \tau_{fast})$ can be obtained by fitting either $G'(\omega)$ with Eq.(4), or $G''(\omega)$ with Eq. (5) or $G(t)$ with Eq.(6). We note that, for each sample, the three procedures give sets of very close numerical values for the fit parameters. We choose in the following to show as numerical results of the fit parameters the average of the values obtained independently with Eqs. (4) and (5). Accordingly, the errors bars shown in the plots (figures 6,7,8) correspond to the amplitude of variations of each fitting parameter measured independently by the fits of $G'$ and $G$. Similarly, the fitting parameters ($G_H = G_\infty$, $\tau_H = \tau$) for naked or hairy micelles shown in the plots as the average of the values extracted from the fits of the data with Eqs. (1, 2).

In Figures 6 and 7, we show the dependence with the amount of copolymers of the fit parameters, the plateau modulus and characteristic time for the fast and slow modes respectively, $G_{fast}$ and $\tau_{fast}$, and $G_{slow}$ and $\tau_{slow}$ respectively for telechelic copolymers. For comparison, the plateau modulus and characteristic time, $G_H$ and $\tau_H$, for hairy wormlike micelles are also plotted.

Independently of the amount of amphiphilic polymers that decorate the micelles, the plateau modulus, $G_H$, is constant and equal to that of a solution of pure WM. By contrast, both $G_{fast}$ and $G_{slow}$ increase as the amount of telechelic polymers increases. The growth of $G_{slow}$ is moderate, since it varies from about 155 Pa (its value for naked micelles) to about 335 Pa for $\beta = 2.5\%$. Inversely, the increase of $G_{fast}$ is steeper with the amount of telechelic polymers. The fast modulus varies linearly with $\beta$ and reaches 1130 Pa for $\beta = 2.5\%$. A linear fit is plotted in figure 6, which shows that $G_{fast}$ extrapolates perfectly to 0 for $\beta = 0$. Hence, these results demonstrate that the slow relaxation process is related to the transient network formed by the wormlike micelles whereas the fast relaxation mode originates from the presence of telechelic polymers.

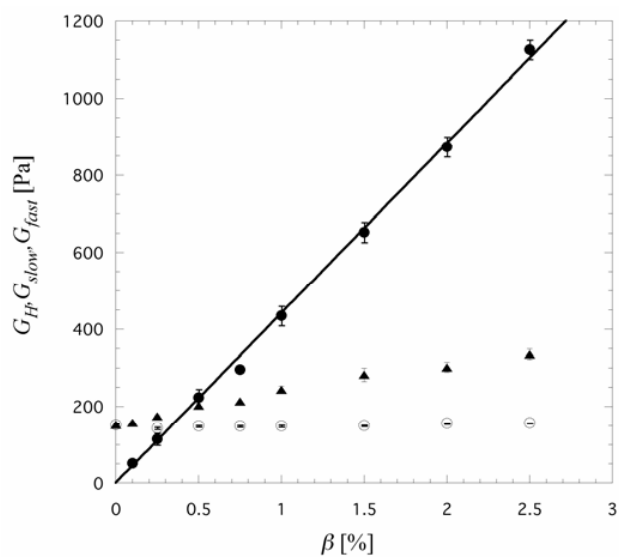

Figure 6: The plateau moduli for hairy WM, $G_H$, (opened circles), and for bridged WM, $G_{slow}$ (closed circles) and $G_{fast}$ (closed triangles) as a function of the stickers of telechelic or amphiphilic polymer over surfactant molar ratio $\beta$, as determined from the fits of the frequency dependence of the loss and storage modulis. The line is a linear fit of $G_{fast}$. The surfactant concentration is $\phi = 9\%$ and the temperature is $T=30°C$.

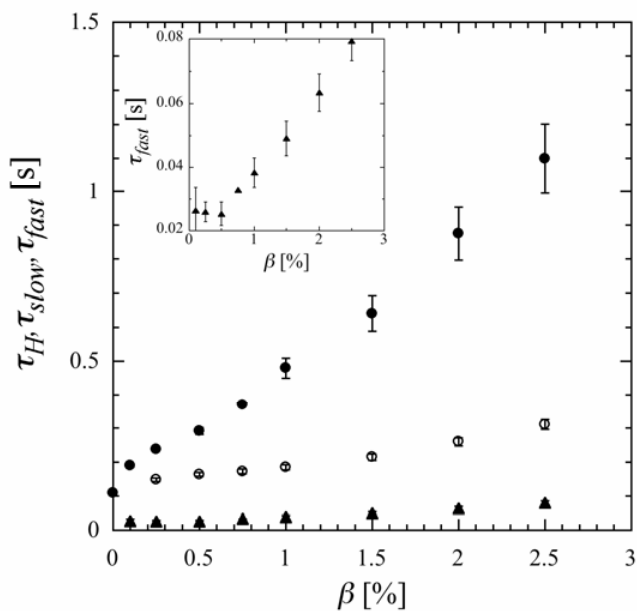

Figure 7: The relaxation times for hairy WM, $\tau_H$, (opened circles), and for bridged WM, $\tau_{slow}$ (closed circles) and $\tau_{fast}$ (closed triangles) as a function of the stickers of telechelic or amphiphilic polymer over surfactant molar ratio $\beta$, as determined from the fits of the frequency dependence of the loss modulus. Inset: zoom up of the $\tau_{fast}$ plot. The surfactant concentration is $\phi = 9\%$ and the temperature is $T=30°C$.

As shown in Figure 7, for bridged wormlike micelles, the fast characteristic time is typically one order of magnitude smaller than the time for naked and decorated micelles and increases linearly with $\beta$. This time, which is directly related to the presence of links between the micelles, is roughly 4 times higher for the largest amount of telechelic polymer investigated ($\beta = 2.5\%$) than for the smallest amounts ($\beta = 0.125 - 0.25\%$). The slow mode is related to the relaxation of the entangled micelles. It nevertheless appears to be strongly dependent on the amount of telechelic polymers present in the material. As shown in Figure 7, the characteristic time of the slow mode, $\tau_{slow}$, increases linearly with $\beta$ and is one order of magnitude higher for a sample with $\beta = 2.5\%$ than for a sample without telechelic polymer. We note that the single characteristic time for hairy wormlike micelles continuously increases with the amount of copolymer as well, with however a much slower growth than that for bridged WM: a three-fold increase is measured, when $\beta$ varies from 0 to 2.5%. This clearly indicates that the relaxation processes of hairy and bridged wormlike micelles are different.

In this section, we have presented experimental data for samples with a constant surfactant concentration ($\phi = 9\%$) and varying copolymer amounts $\beta$ (vertical line in the phase-diagram figure 1). Two other series of samples (horizontal lines in the phase-diagram figure 1) have also been investigated, for which the surfactant concentrations vary. The viscoelasticity of these samples display the same features as that of the samples described here: the viscoelasticity of bridged wormlike micelles is characterized by two Maxwell modes, whereas that of hairy micelles can be well described by a single Maxwell mode.

## 3 Discussion

**Bridged wormlike micelles : a double transient network**

Our experimental data clearly show that mixtures of wormlike micelles and telechelic polymers consist of two interpenetrated transient networks that relax independently.

The first network is the substrate network of entangled wormlike micelles which exhibits a Maxwell relaxation behavior, slower than the second Maxwell relaxation. Both the relaxation time, $\tau_{slow}$, and the shear modulus, $G_{slow}$, are however affected by the presence of the telechelic chains since they increase continuously from the numerical values obtained for naked micelles, with increasing amounts of telechelic polymers, $\beta$ (Figures 6 and 7). Entangled wormlike micelles nevertheless preserve their simple Maxwell behavior, whose microscopic origin can be explained satisfyingly by Cates theory [Cates, (1987)], despite the fact that they are linked by polymeric junctions.

On the other hand, we expect the second transient network to be related to the bridges that are interconnected through the substrate network. In addition, since the substrate network of entangled micelles has a relaxation time one order of magnitude larger that the relaxation time of the network of telechelic chains ($\tau_{fast} << \tau_{slow}$), the micellar network can be considered as a permanent network for the telechelic chains.

We compare the zero shear viscosity, $\eta(\dot{\gamma} \rightarrow 0)$, measured in a steady shear measurement to the viscosity extracted from dynamic measurements, $\eta_{dyn}$, that is the sum of the viscosities of each Maxwell fluid of the blend:

$$\eta_{dyn} = \int_0^\infty G(t)dt = G_{slow}\tau_{slow} + G_{fast}\tau_{fast} = \eta_{slow} + \eta_{fast} \qquad (7)$$

Both $\eta(\dot{\gamma} \rightarrow 0)$ and $\eta_{dyn}$ are plotted in Figure 3. The excellent collapse of the two sets of data is an additional check that bridged wormlike micelles can be considered as a blend of two Maxwell fluids.

Finally, we note that, to the best or our knowledge, experimental realizations of viscoelastic fluids exhibiting only two well defined Maxwell modes are very rare. Previous realizations of such a blend include a water solution of a 50:50 mixture of two telechelic polymers with hydrophobic ends of 12 and 20 carbon units (with however fits of $G''(\omega)$ of lower quality than ours) [Annable, et al., (1993)], and a water solution of telechelic polymers [Ng (2000)] Ng et al claimed that the life time of the hydrophobe in the micelle and the relaxation time of the network can be distinguished leading to the existence of two modes.

**Fast mode of bridged wormlike micelles**

*Plateau modulus $G_{fast}$*

The fast Maxwell mode is related to the network formed by the bridges. The Tanaka and Edwards theory of transient networks [Tanaka and Edwards, (1992)] predicts a linear increase of the plateau modulus with the molar concentration $\mu_b$ of bridges:

$$G_\infty = \mu_b RT \qquad (8)$$

where $R$ is the constant of ideal gases and $T$ is the temperature.

We expect the molar concentration of bridges to be proportional to the molar concentration of telechelic chains, $\mu$, with a proportionality factor, $f_b$, that represents the fraction of bridges: $f_b$=[bridges]/([bridges]+[loops]). On the other hand, the molar concentration of telechelic polymers is proportional to the product $\beta\phi$: $\mu = \alpha\beta\phi$. We calculate that the proportionality factor, $\alpha$, which depends on the molecular weight of the surfactants (CpCl and NaSal) and on the molar ratio of the two surfactants, is equal to $1.19 \times 10^3$ mol/m$^3$.

For samples with constant surfactant concentration, we expect that the fraction of bridges, $f_b$, is constant, since the correlation length of the micellar solution, $\xi$, which corresponds to the mean distance between two neighboring cylinders is constant as well. This is probably not the case for samples comprising different surfactant concentrations since $\xi \propto \phi^{-3/4}$. Hence, for samples with $\phi$=9% and variable $\beta$, one predicts that $G_{fast}$ varies affinely with $\beta$, as observed experimentally (Figure 6).

Figure 8 shows the variation of $G_{fast}$ as a function of $\beta\phi$ for the three series of samples we have investigated (see Figure 2). Interestingly, one observes that all data collapse on a single master curve. This shows that the fraction of bridges is constant for all samples, hence even for samples with different surfactant concentration. From the slope of the fit and according to $G_{fast} = f_b(\alpha RT)\phi\beta$, we can measure the fraction of bridges. We find : $f_b \approx 0.16$. This value is of the correct order of magnitude. Indeed, when the average distance between the micelles is of the order of the radius of gyration of the POE block, or samller than, one expects to have as many loops as bridges, thus $f_b$ =0.5 [Porte, et al., (2006)]. Note that, our evaluation of $f_b$ is certainly underestimated. Eq.8 is the prediction for an affine network, while a phantom model seems to be more appropriate, since, in our experimental configuration, the hydrophobic stickers are free to diffuse along the cylinders and fluctuate around their average positions. For a phantom network the elastic plateau is also proportional to $\mu_b RT$ but with a proportionality factor smaller than 1 and which depends on the functionality of the junction [Staverman, (1982)].

A striking result is that it appears that $f_b$ is constant even for samples with small surfactant concentration ($\phi$ is as low as 1%). This has to be contrasted with binary solutions of telechelic polymers, for which the fraction of bridges decreases with decreasing polymer concentration, as shown by Annable, *et al.*, (1993). This finding reflects the fact that bridges configurations are entropically disadvantaged with respect to loops configurations when the mean distance between objects increases. However one has to keep in mind that in our case, stickers can slide along the fluid surfactant micelles, and so they can accumulate around the entanglements of the cylinders in order to relax their conformational entropic penalty with respect to loops configuration that do not contribute to the elasticity. Note that this may also explain the fact that, for low $\beta$ (0.25%), one-phase samples are obtained for all surfactant concentrations, while at higher $\beta$, samples phase-separate for surfactant concentration below a critical value.

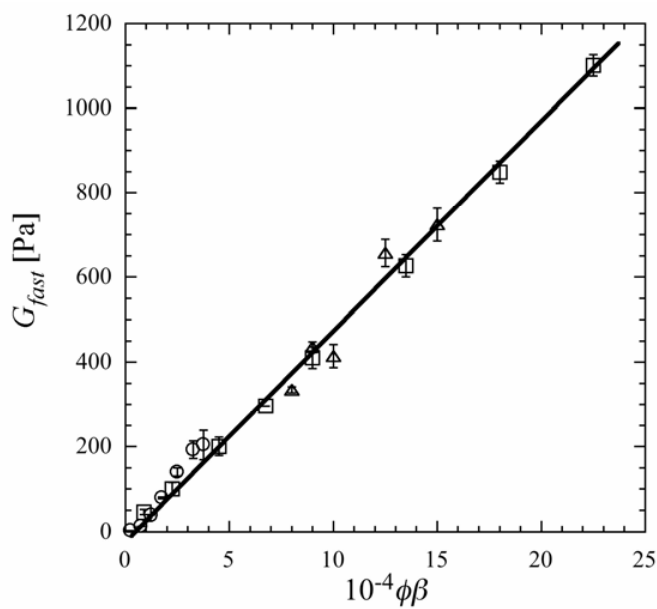

Figure 8 Plateau modulus of the fast mode $G_{fast}$ for bridged WM at $T=30°C$ as a function of the concentration of telechelic polymers $\beta\phi$ for three series of samples: $\phi =$ 9%, varying $\beta$ (squares), varying $\phi$ and $\beta = 0.25\%$ (circles) and $\beta = 1\%$ (triangles). The solid line is the best linear fit of $G_{fast}$.

Finally, it is worth emphasizing one important and original property of the secondary network of telechelic chains: for our composite material, the telechelic chains form a transient network at any arbitrary low polymer concentration (as shown in figure 8) in marked contrast with the more classical situation of a binary solutions of telechelic chains or other similar systems like bridged microemulsions [Filali, *et al.*, (2001)], that form a gel above some percolation concentration, and behave as simple fluids below this concentration. The reason is that entangled wormlike micelles form a substrate network for active telechelic chains which can be considered as permanent for the second network since $\tau_{slow} \gg \tau_{fast}$.

Consequently any stress applied to a single bridge will be transmitted in the entire material through the substrate network.

*Relaxation time $\tau_{fast}$*

The inset of figure 7 shows that for a constant volume fraction of micelles, the fast relaxation time increases linearly with $\beta$, for $\beta$ larger than 0.5%. Below this value, the $\tau_{fast}$ data are of the same order of magnitude as the uncertainty of the measurements. This property seems to be in contradiction with the framework of the transient network theory for which the relaxation time is the characteristic time of the random disengagement process from the hydrophobic core of the micelles of a single sticker ; $\tau_{fast}$ is thus expected to be independent of the number of junction points, and is related to the probability, $w(t)$, that a bridge breaks during the interval $[t, t +\Delta t]$. $w(t)$ is given by the exponential distribution:

$$w(t) = \frac{1}{\tau_M} e^{-t/\tau_M} \qquad (9)$$

where

$$\tau_M = \tau_0 e^{E_m/kT} \qquad (10)$$

In Eq.10, $\tau_0$ is a characteristic time of thermal vibration estimated by Tanaka and Edwards to be of order of 0.1ns and $E_m \approx n_{CH_2} kT$ is a potential barrier to disengagement, proportional to the number of methylene group per sticker $n_{CH_2}$ [Annable, *et al.*, (1993)].

In our case, however, the topology of the network is such that, two neighboring entangled wormlike micelles are connected by $n_b$ bridges in parallel. Each bridge relaxes randomly and independently according to Equation (9), and by definition of $\beta$, $n_b \propto \beta$. Accordingly, the probability that $n_b$ bridges break in the interval $[0,t]$ is given by a Poisson distribution: $P_{n_b}(t) = \frac{(t/\tau_M)^{n_b}}{n_b!} e^{-t/\tau_M}$. For $n_b \gg 1$, the Poisson distribution $P_{n_b}(t)$ has a sharp maximum at its mean value $\langle \tau \rangle = n_b \tau_M \propto \beta \tau_M$. The rupture of all the links between two neighboring cylinders defines the fast relaxation time, so that $\langle \tau \rangle \equiv \tau_{fast} \propto \beta$ for high enough $\beta$, as observed experimentally.

Finally, we have checked that the Tanaka and Edwards transient network theory adequately describes the temperature dependence of the fast relaxation. As shown in Figure 9, an Arrhenius law is measured experimentally, as predicted by Eq. (10), for samples with $\phi = 9\%$ and various telechelic concentrations $\beta$. We find for the activation energy $E_m(\beta = 0.25\%) = 67.2\ k_BT$, $E_m(\beta = 1.5\%) = 77.9\ k_BT$ and $E_m(\beta = 2\%) = 78.9\ k_BT$. These values are much larger than the theoretically expected value (18 $k_BT$) and slightly depend on the amount of copolymer. These two features probably point out a mechanism more complex than the one presented before, but that remains to be elucidated.

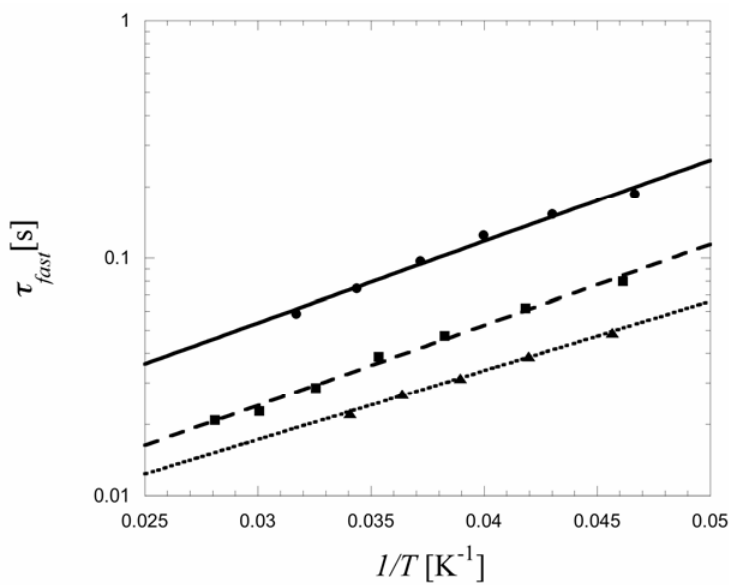

Figure 9 The fast relaxation time $\tau_{fast}$ of bridged wormlike micelles at constant volume fraction of surfactant $\phi = 9\%$ as a function of the inverse of temperature $(1/T)$ for three samples : $\beta = 0.25\%$ (triangles), $\beta = 1.5\%$ (squares), $\beta = 2\%$ (circles). The lines are the best fits of Equation (10). The temperature varies between 21°C and 31 °C.

**Slow mode of the bridged wormlike micelles**

*Relaxation time $\tau_{slow}$*

Figure 7 shows that the slow relaxation time increases linearly with $\beta$ for $\beta > 0.25$, indicating that bridges slows down the relaxation process of the entangled wormlike micelles. Moreover, we note that this slowing down is much more important that for hairy wormlike micelles with the same composition. This shows that the origin of this slowing down is due to the presence of bridges between the entangled cylinders. Indeed bridged wormlike micelles present strong analogies with entangled sticky chains (entangled melt or solutions of hydrophobically modified polymers ), despite the fact that the cylinders are breakable. Leibler*, et al.*, (1991) have studied theoretically the dynamics of entangled networks made up of linear chains with many temporary cross-links. They have shown that at times shorter than the lifetime of a cross-link such networks behave as elastic rubbers (gels). On longer time scales the successive breaking of only a few cross-links allows the chain to diffuse along its confining tube. The motion of a chain in this hindered reptation model called "sticky reptation" is controlled by the concentration and lifetime of tie points, and the reptation time scales as $S^2$, where $S>>1$ is the mean number of stickers per chain. So, even if a theory of the sticky reptation of breakable chains remains still to be done, we can use the simple following argument. In our experimental system, $S$ can be identified to the number of bridges per surfactant cylinders, so that $S \propto \beta$, and for high enough $\beta$, one expects $\tau_{rep}(\beta) \propto \beta^2$. Assuming now that the breaking time of surfactant cylinders is not modified by the presence of telechelic chains and using the Cates prediction for the relaxation time of entangled wormlike micelles, one expects that $\tau_{slow}(\beta) = \sqrt{\tau_{rep}(\beta)\tau_{break}} \propto \beta$, in agreement with the experimental data.

*Plateau modulus $G_{slow}$*

Surprisingly, we observe that the plateau modulus of the slow mode, related to the relaxation of the entangled micelles increases with $\beta$ (see Figure 6). This is an unexpected result. Indeed, the sticky reptation model of Leibler, *et al.*, (1991) predicts that the plateau modulus is not modified by the tie points. The authors however mention that the effective tube diameter may be decreased by the presence of the temporary cross-links but do not develop this point. Dynamical scaling analysis [Kavassalis and Noolandi, (1989)] computer simulations [Muller, *et al.*, (2000)] and experiments [Colby, *et al.*, (1991)] have shown that the elastic modulus of a semi-dilute solution of wormlike micelles (or conventional linear polymers) is proportional to the density of entanglements:

$$G_0 = \frac{kT}{N_e \xi^3} \tag{11}$$

In Eq. (11) $\xi$ is the blob size: $\xi = l_0 \phi^{-3/4}$, where $l_0 = l_K^{1/4} v^{-1/4} s^{3/4}$. The Kuhn length $l_K$ of the chains is equal to twice the persistence length $l_p$, $s$ is the cross-section of the micelles, $v$ the excluded volume and $N_e$ is the number of Kuhn monomers per entanglement strand in the melt state. The facts that the plateau modulus of hairy wormlike micelles is not modified by the grafting chains allow one to infer that the blob size of wormlike entangled micelles is not significantly modified by the presence of telechelic chains. By a process of elimination, we conclude that the presence of bridges linking the cylinders may decreases $N_e$ even if the life time of these links is much shorter than the relaxation time of the micelles; but we cannot provide any explanation for this behavior presently.

**Hairy wormlike micelles**

We now briefly discuss the linear rheological of bridged wormlike micelles. The most important result is that the viscoelastic properties of entangled micelles are not markedly affected by the presence of grafted layers of polymers which decorate the cylinders. In particularly hairy wormlike micelles behave as a simple Maxwell fluid with the same plateau modulus $G_H$ as for naked micelles at the same concentration, as shown clearly in Figure 6. This important result demonstrates, according to equation (11), that the persistence length, as well as the excluded volume are insensitive to the amount of grafted polymers. A different behavior has been observed by Massiera, *et al.*, (2002) with a different amphiphilic copolymer (a triblock PEO-PPO-PEO copolymer). In this latter case the hydrophobic block had a very different chemical nature of that of the hydrophobic tail of the CPCl surfactant, and played the role of a cosurfactant, which can deeply modify the size distribution or the topology of the cylinders.. The choice of a hydrophobic sticker for the amphiphilic copolymer (as well as for the telechelic polymer) with a chemical composition very close to that of the surfactant tail, as in the present system, shows that grafted polymers layers by themselves do not modify the elastic modulus of the transient network.

As shown in Figure 7, the single characteristic time for hairy wormlike micelles continuously but slowly increases with the amount of copolymers (much more slowly than the characteristic time of the slow mode of bridged wormlike micelles). Qualitatively, this effect could be attributed to a enhancement of the friction coefficient of a cylindrical micelle along its reptation tube. In the mushroom regime, this enhancement should be proportional to the grafting density of polymer chains, so that the effective local solvent viscosity should increase linearly with $\beta$: $\eta_{loc}(\beta) \approx \eta_{loc}(0)(1+\lambda\beta)$. Consequently the Rouse time of a cylindrical micelle should vary as $\tau_R(\beta) \approx \tau_R(0)(1+\lambda\beta)$. The slight deviation (increasing with $\beta$) from a perfect semi-circle of the Cole-Cole plot for hairy wormlike micelles in the high frequency domain observed in the inset of Figure 4 are compatible with this assumption This deviation has been attributed to a crossover towards the breathing regime in the relaxation process of wormlike micelles: the higher is the Rouse time, the smaller is the crossover in the frequency space. Finally, since the reptation time is itself proportional to the Rouse time, and assuming again that the breaking time of the micelles is independent of $\beta$, one expects from Cates theory $\tau_H(\beta) \approx \tau_H(0)\sqrt{1+\lambda\beta}$. This simple model is much too crude to expect a quantitative agreement with the experimental data, but is in qualitative agreement with them.

## Conclusions

Solutions of entangled wormlike micelles on the one hand and of telechelic polymers on the other hand are the paradigms of transient self-assembled networks that behave as perfect Maxwell fluids. By mixing them, we have designed a double transient network of WM bridged by telechelic polymers. We have shown that this soft composite material behaves as a near perfect blend of two Maxwell fluids that relax independently. One Maxwell fluid is associated with a network of WM and the other one with a network of telechelic chains. Accordingly, the zero shear modulus of the material has been found to be equal to the sum of the zero shear modulus of the surfactant micelles, which originates from the entanglements between the WM, and the zero shear modulus of the sliding junctions due to telechelic chains. Interestingly, this result confirms a quite old conjecture of polymer physics proposed first for entangled cross-linked polymers melts, and recently rigorously established by Oberdisse, *et al.*, (2006) using Brownian dynamic simulations, i.e., the additivity assumption of slip- and crosslink contribution to the strain response for sufficiently long chains with two or more entanglements.

The slow mode is associated with the transient network of entangled wormlike micelles, whose relaxation can be well understood in the framework of the breakable reptation theory of Cates, (1987). However, the presence of transient junctions, i.e. the telechelic bridges, between the micelles slows downs the relaxation process in agreement with the sticky reptation model of Leibler, *et al.*, (1991). We have also measured a linear dependence of the zero shear modulus with the amount of telechelic chains, that remains to be understood.

The transient network formed by the telechelic chains is well described by the transient network theory of Tanaka and Edwards, (1992): the elastic modulus is proportional to the density of active chains, and the relaxation time shows an Arrhenius law for the temperature dependence. It increases linearly with the amount of telechelic chains per wormlike micelle due to the fact that many bridges link in parallel two neighboring cylinders. Interestingly, this transient network exists at arbitrary low polymer concentration without any percolation transition, contrary to all other experimental realizations of transient networks made from telechelic polymers. The reason is that this network is supported by the substrate network of entangled micelles which has a much more longer relaxation time.

In parallel, we have investigated entangled solution of wormlike micelle decorated by amphiphilic polymers, whose visco-elasticty has been found to be characterized by a simple Maxwell fluid with characteristic time and plateau modulus very close to those of naked WM. This proves unambiguously that the existence of a second Maxwell mode in entangled micelles mixed with telechelic polymers has to be attributed to bridges.

Non linear rheological properties of the double transient network are out scope of this work; they will be the object of future investigations.

## Acknowledgments

We thank T. Phou and R. Aznar for the polymer synthesis, and G. Porte and S. Mora for fruitful discussions. K. Nakaya thanks the Région Languedoc-Roussillon for partial financial support. We thank M. Imai for having promoted this collaboration. This project has been supported in part by the European Commission under the sixth Framework Programme through Integrating and Strengthening the European research Area, Contract SoftComp, NoE/NMP3-CT-2004-502235, and by the Agence Nationale de la Recherche, under the Programme blanc, CDS4, contrat ANR-06-BLAN-0097-01 "Tailored Transient Self Assembled Networks".